\documentclass[pre,a4paper,twocolumn,superscriptaddress,amsmath,amssymb,preprintnumbers,showpacs,nofootinbib,showkeys,tightenlines,floatfix]{revtex4}
\usepackage{latexsym}
\usepackage{epsfig}
\usepackage{epstopdf}
\usepackage{amssymb}
\usepackage{amsmath}
\usepackage{color}
\usepackage{graphicx}

\newcommand{\bea}{\begin{eqnarray}}
\newcommand{\eea}{\end{eqnarray}}

\newcommand{\be}{\begin{equation}}
\newcommand{\ee}{\end{equation}}

\definecolor{Mygreen}{cmyk}{0,1.0,0,0}
\definecolor{MyOrange}{cmyk}{0.5,1.0,0,0}

\newcommand{\bi}{\begin{itemize}}
\newcommand{\ei}{\end{itemize}}

\begin{document}
\title{An equivalence between a Maximum Caliber analysis of two-state kinetics and the Ising model}
\author{Sarah Marzen}
\affiliation{California Institute of Technology, Pasadena CA}
\author{David Wu}
\affiliation{California Institute of Technology, Pasadena CA}
\author{Mandar Inamdar}
\affiliation{Indian Institute of Technology, Mumbai, India}
\author{Rob Phillips}
\email{phillips@pboc.caltech.edu}
\affiliation{California Institute of Technology, Pasadena CA}

\date{\today}

\begin{abstract}
Application of the information-theoretic Maximum Caliber principle to the microtrajectories of a two-state system shows that the determination of key dynamical quantities can be mapped onto the evaluation of properties of the 1-D Ising model.  The strategy described here is equivalent to an earlier Maximum Caliber formulation of the two-state problem, but reveals a different way of imposing the constraints which determine the probability distribution of allowed microtrajectories.  The theoretical calculations of second moments, covariances, and correlation times that are obtained from Maximum Caliber agree well with simulated data of a particle diffusing on a double Gaussian surface, as well as with recent experiments on a particle trapped by a dual-well optical trap.  The formalism reveals a new relationship between the average occupancy of the two states  of the system, the average number of transitions between the two states that the system undergoes, Markov transition probabilities, and the discretization time step.  In addition, Maxwell-like relations imply
how measurements on one potential landscape can be used to make predictions 
about the dynamics on a different potential landscape, independent of further experiment.
\end{abstract}

\maketitle

\section{Introduction}
\label{section:Introduction}

The Maximum Entropy (MaxEnt) formalism provides an alternative to more conventional statistical mechanics approaches by offering a convenient jumping off point for thinking about problems that are not within the purview of equilibrium statistical mechanics.  For non-equilibrium systems, the corresponding idea is called the Principle of Maximum Caliber (MaxCal), in which we attempt to determine the probability that a particle travels on a particular space-time path rather than its probability of occupying a certain energy level \cite{Jaynes}.  One way to think about these information-theoretic approaches is that the probability distributions generated by MaxEnt and MaxCal are the least-biased inferences about the distributions that can be made on the limited information that is in hand; i.e., by maximizing the Shannon entropy over the collection of possible
outcomes for the system of interest \cite{Shannon}.  MaxEnt has been used for several decades to model everything from neural firing \cite{Bialek:neurons} to climate change \cite{Karmeshu} and many applications between.  MaxCal has been subjected to a few applications including analysis of the dynamics of a two-state system \cite{Gerhard, Wuetal} and to examine the origin of phenomenological transport laws including Fick's law of diffusion \cite{Seitaridou} and fluctuation theorems \cite{RDewar}.

 
The MaxCal approach is relatively unexplored and has been set forth as a very general method of analyzing nonequilibrium systems \cite{Jaynes, Jaynes2, Jaynes3}.
By applying MaxCal to experimentally realizable systems, we hope that we can not only make useful predictions of the behavior of concrete experimental systems, but also help identify and highlight the outstanding theoretical questions that impede progress towards practical use of MaxCal.  One such system that we find appealing is the dynamics of a two-state system, partly because of its simplicity and because of the opportunity to manipulate such systems experimentally.  To that end, in this paper, we analyze the dynamics of a two-state system, cognizant that it is only one concrete application of the MaxCal approach and that much remains to be done to demonstrate MaxCal's usefulness more generally.

MaxEnt as used in equilibrium statistical mechanics and MaxCal exploit almost entirely the same theoretical prescription, except that MaxEnt as used in equilibrium statistical mechanics identifies the \textit{entropy}-maximizing probability distribution of particle occupancy over energy states, whereas MaxCal identifies the \textit{caliber}-maximizing probability distribution of particle occupancy over possible space-time paths (``microtrajectories'').  Caliber is an entropy-like quantity that has the same form as the Shannon entropy, ${\cal C} = \sum_{\Gamma} p_{\Gamma} \ln p_{\Gamma}+\sum_{i}\lambda_{i} \langle A_{i} \rangle_{\Gamma}$.    In this equation, ${\cal C}$ is the caliber, $p_{\Gamma}$ denotes the probability of a given microtrajectory  $\Gamma$, and each $\langle A_{i}\rangle_{\Gamma}$ is a macroscopic observable which serves as a constraint within
the information-theoretic approach.  For convenience, we will denote an average over the microstates $\langle A_{i} \rangle_{\Gamma}$ as $\langle A_i\rangle$.  The constants $\lambda_{i}$ are Lagrange multipliers corresponding to those constraints.  Maximizing the caliber yields a full probability distribution for all the allowed microtrajectories.

In this paper, we analyze the dynamics of the two-state system using a MaxCal model and derive new insights into MaxCal and the discretized two-state system in the process.  The organization of this paper is as follows.  Background on MaxCal is presented in Section \ref{section:Background}.  In Section \ref{section:StatisticalDynamics}, we show that the dynamics of a two-state system are mathematically equivalent to the 1-D Ising model, and that there are different but equivalent ways of picking the MaxCal constraints.  In Section \ref{section:Test}, we use simulation to investigate the relationship between the MaxCal model for the two-state system proposed here and a previously proposed MaxCal model for the two-state system \cite{Gerhard, Wuetal}.

\section{Background}
\label{section:Background}

As noted above, caliber is an entropy-like quantity that increases as the relative probabilities of the microtrajectories becomes less predictable.  As a result of maximizing the caliber, MaxCal asserts that our best guess as to the probability of observing a particular microtrajectory $\Gamma$ is
\begin{equation}
p_{\Gamma}=\frac{1}{Z}\exp(\sum_{i} \lambda_i A_{i,\Gamma}),
\label{eq:MaxCalProbGeneral}
\end{equation}
where
\begin{equation}
Z=\sum_{\Gamma}\exp\left(\sum_i \lambda_i A_{i,\Gamma}\right)
\label{eq:MaxCalPartFnctnGeneral}
\end{equation}
is a normalization factor dubbed by Jaynes as the ``partition functional'' and
the  $A_{i,\Gamma}$ are values of macroscopic observables $A_i$ along the microtrajectory labeled $\Gamma$.  $\lambda_i$ are the corresponding Lagrange multipliers that can be calculated from $\langle A_{i}\rangle=\frac{\partial\ln Z}{\partial\lambda_i}$  \cite{Jaynes}.

A distinguishing feature of MaxCal-derived formulas of nonequilibrium statistical mechanics as compared to more conventional nonequilibrium statistical mechanics formulas, e.g. the Jarzynski equality \cite{Jarzynski}, is that MaxCal solves an inverse problem, in which we infer microscopic properties based on macroscopic observations.
The MaxCal distribution described in Equation \ref{eq:MaxCalProbGeneral} is typically generated on
the basis of  knowing only a small number of macroscopic observables $\langle A_i\rangle$ and predicts the probability of traveling any one of the many possible microtrajectories $\Gamma$.  For example, in the two-state systems simulated in this paper which each have $10^5$ time steps, only two macroscopic observables are used to generate predictions for the probability of traversing the entirety of the  $2^{10^5}$ possible microtrajectories.  However, since the average probability of observing a particular microtrajectory is typically very small, it is difficult to obtain accurate experimental estimates of $p_{\Gamma}$ itself to which we can compare MaxCal predictions.  Instead, to test the accuracy of a MaxCal model, we can predict higher order moments of the joint probability distribution $p(A_1,A_2,...,A_n)$.  These higher moments are more robust than individual microtrajectory probabilities $p_{\Gamma}$ to the depth of our experimental investigation of the system.  Again, note that these higher moments are {\it not} used in any way to determine the probability distribution $p(A_1,...,A_n)$ itself.   Rather, they  are predictions about more nuanced features of the two-state system than those simple average quantities used to establish the distribution given in Equation \ref{eq:MaxCalProbGeneral}.

\section{Statistical Dynamics of the Two-State System Using Caliber}
\label{section:StatisticalDynamics}

The dynamics of the two-state system can be completely characterized by  rate equations in terms of the constants $k_A$ and $k_B$, where $k_A$ and $k_B$ are the intrinsic rate constants for transition from state A to state B and vice versa \cite{Gardiner}.  Indeed, one of the interesting outcomes of the present paper is a simple relation between the conventional rate constants and the Lagrange multipliers that are central to the MaxCal description of the same system.

\begin{figure}[h]
	\centering
  \includegraphics[width=8.6cm]{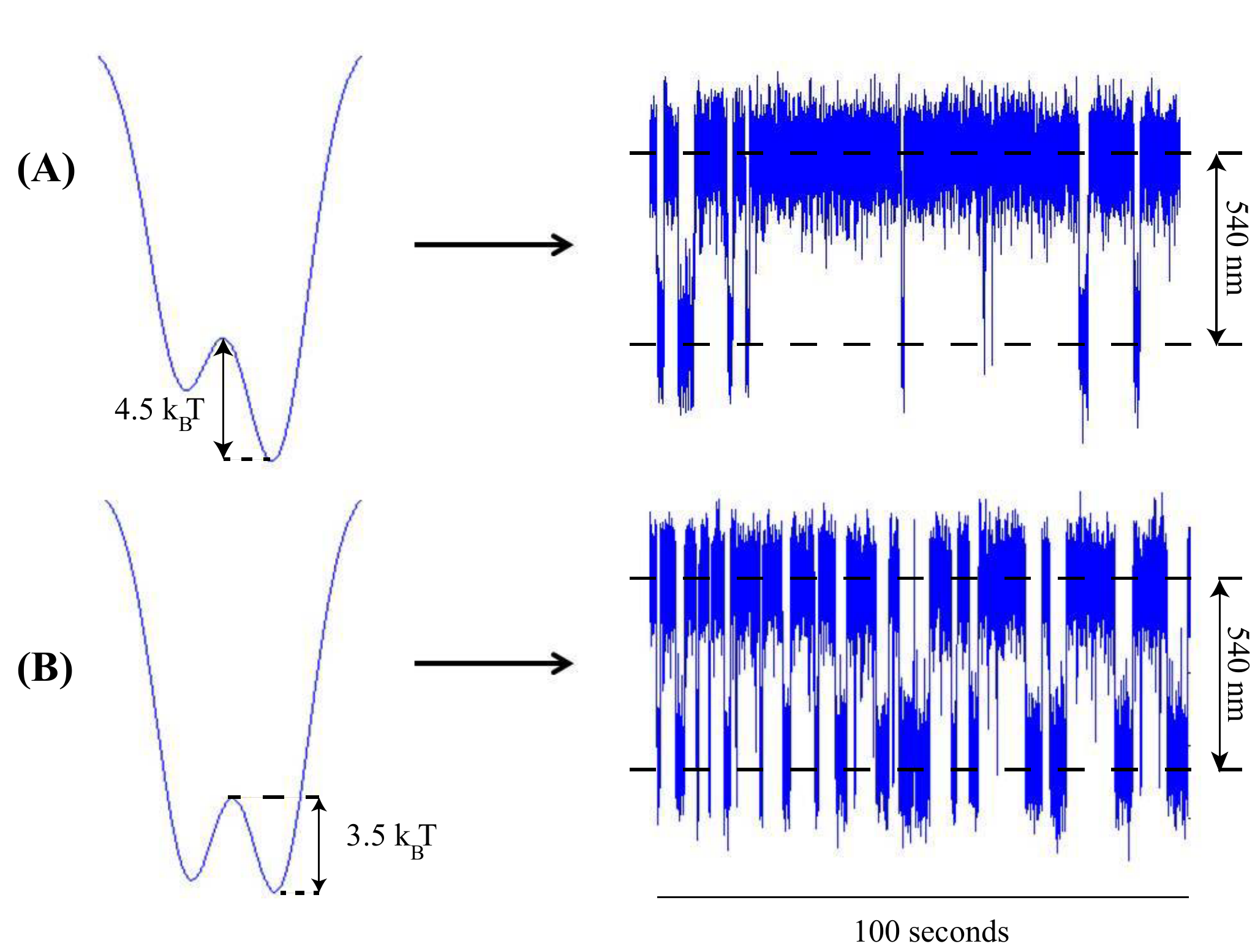}
  \caption{\textbf{Representative simulated trajectories on two different two-well potential energy landscapes.} A simulation of a particle diffusing on a double Gaussian potential surface generates trajectories of the particle diffusing on the energy landscape according to Equation (\ref{eqSimulation}).  The traces show the raw data.  States are assigned after boxcar filtering and threshold finding.  (A) Potential well number 1.  See Table \ref{table:parameters2}.  The state of lower potential energy is more frequently populated; there is a high barrier, with infrequent transitions.  (B) Potential well number 8.  See Table \ref{table:parameters2}.  The upper and lower states are populated roughly equally.}
  \label{fig:Data}
\end{figure}

To make the dynamics of the two-state system amenable to simple MaxCal analysis, let the MaxCal trajectory $\Gamma$ be a string of $1$'s and $-1$'s describing the state of the system as a function of time.  More explicitly, suppose that our two-state system is a particle jumping between wells on a double-well potential energy landscape, examples of which are shown in Figure \ref{fig:Data}.  Let $\sigma_{\Gamma}(t)$ represent the system's state at time $t$.  If the particle is in the left potential well at time $t$, then we say it is in state $A$ and that $\sigma_{\Gamma}(t)=1$; if the particle is in the right potential well at time $t$, then we say it is in state $B$ and $\sigma_{\Gamma}(t)=-1$.  Thus, a particle's trajectory $\Gamma$ is equivalent to a series of $N$ $1$'s and $-1$'s, for which the particle's state is sampled every $\Delta t$ and the total time spent observing the particles of interest is $N \Delta t$.

This discretized picture of the two-state process will lead to accurate calculations only if our time step $\Delta t$ is small enough to capture nearly all state transitions \cite{Hawkes1}.  In other words, in the language of rate constants, we require that the average waiting time in both states is much larger than the time step $\Delta t$.  For Markov systems, the waiting time in a particular state $t$ follows an exponential distribution with a mean of $\langle t\rangle=\frac{1}{k}$ \cite{Gardiner}.  Thus, here, we require $\Delta t<<\frac{1}{k_A},~\frac{1}{k_B}$, or identically $k_A\Delta t,~k_B\Delta t<<1$ as described previously  \cite{Wuetal}.  If this heuristic condition is satisfied, missed transitions are negligible.  For the two-well potential landscapes considered here, waiting times range from $0.5$ to $5$ s, whereas $\Delta t=1$ ms.

\subsection{The dynamics of the two-state system is equivalent to the one-dimensional Ising model}
\label{section:IsingModel}

There are many different ways that one can imagine characterizing the trajectories realized by the two-state system as it sweeps out a telegraph signal \cite{Gardiner}.  The particular choice that is made has implications in turn for the nature of the MaxCal constraints used to describe the system.  
However, it is still unclear to us what general criteria must be used when choosing constraints $\langle A_i\rangle$ for some
specific new problem.

For the two-state system considered here, we focus on two distinct constraints: the average number of times that the particle switches states over the course of a microtrajectory ($\langle N_s\rangle$), and the ``aggregate state'' of the particle ($\langle i\rangle$).  We define the aggregate state mathematically using the state variable $\sigma(t)$, which was defined earlier in Section \ref{section:StatisticalDynamics}, as $i_{\Gamma}=\sum_{t=1}^N \sigma_{\Gamma}(t)$, so that
\begin{equation}
\langle i\rangle=\sum_{\Gamma} p_{\Gamma} \left(\sum_{t=1}^N \sigma_{\Gamma}(t)\right),
\label{eq:TotalState}
\end{equation}
where we obtain the average by summing over {\it all} the possible microtrajectories
of the system (labeled by $\Gamma$), each weighted by its probability $p_{\Gamma}$.
Intuitively, the ``average state'' of the particle ($\frac{\langle i\rangle}{N}$) corresponds to a measure of where the particle is most frequently found.  If the particle is frequently found in state A, then $\langle i\rangle\sim N$; if the particle is frequently found in state B, then $\langle i\rangle\sim -N$; and if the particle spends its time equally between the two states, then $\langle i\rangle\sim 0$.  In terms of experimental quantities, the average state is analogous to the average current if we think of the telegraph signal as the current passing through an ion channel in a single-channel measurement \cite{PBoC}.

With these two constraints and the MaxCal formalism described above, we find that
the discretized version of the partition functional takes the simple form
\begin{equation}
Z=\sum_{\Gamma}\exp(\alpha N_{s,\Gamma} +h i_{\Gamma}),
\label{eqZ1}
\end{equation}
where $\alpha$ and $h$ are the Lagrange multipliers 
associated with our two constraints.  The probability of observing a particular microtrajectory is accordingly
\begin{equation}
p_{\Gamma}=\frac{1}{Z}\exp(\alpha N_{s,\Gamma} +h i_{\Gamma}).
\label{eqProbTraj}
\end{equation}
More precisely, this expression provides the probability of observing a particular microtrajectory $\Gamma$ with aggregate state $i_{\Gamma}$ and number of switches $N_{s,\Gamma}$, where the subscript $\Gamma$ labels the trajectory of interest.

To cast this expression in a more useful way,
we notice that $\langle N_s\rangle $ can be expressed in terms of $\sigma_{\Gamma}(t)$ using the quantity $(\frac{\sigma_{\Gamma}(t+1)-\sigma_{\Gamma}(t)}{2})^2$, which compares the state of the system at two
subsequent instants in time and is $1$ when the particle switches states and $0$ otherwise.  Then, $\langle N_s\rangle=\sum_{\Gamma}p_{\Gamma}(\sum_{t} (\frac{\sigma_{\Gamma}(t+1)-\sigma_{\Gamma}(t)}{2})^2)$ and from combining this fact with Equation \ref{eq:TotalState}, Equation \ref{eqZ1} can be rewritten as
\begin{eqnarray}
Z &=& \sum_{\Gamma} \exp(\frac{\alpha}{2} \left(N-\sum_{t=\Delta t}^{(N-1)\Delta t} \sigma_{\Gamma}(t+\Delta t)\sigma_{\Gamma}(t) \right)\nonumber \\
&& +h \sum_{t=\Delta t}^{N\Delta t} \sigma_{\Gamma}(t) ) \\
&=& \exp\left(\frac{\alpha}{2}N\right) \sum_{\Gamma} \exp(-\frac{\alpha}{2} \sum_{t=\Delta t}^{(N-1)\Delta t}\sigma_{\Gamma}(t+\Delta t)\sigma_{\Gamma}(t) \nonumber \\
&& +h \sum_{t=\Delta t}^{N\Delta t} \sigma_{\Gamma}(t) ).
\end{eqnarray}
Let $J=-\frac{\alpha}{2}$.  Then
\begin{equation}
Z=\exp(-JN)\sum_{\Gamma}\exp(J\sum_t \sigma_{\Gamma}(t)\sigma_{\Gamma}(t+\Delta t)+h\sum_t\sigma_{\Gamma}(t)).
\label{eqZ2}
\end{equation}

The partition function in Equation \ref{eqZ2} looks like the partition function for the 1-D Ising model.  In this context, a spin at position $x$ corresponds to the particle's state at time $t$; the lattice spacing $\Delta x$ corresponds to the time step $\Delta t$; $h$ is analogous to the applied magnetic field; and $J$ is the analog of the coupling constant between spins.  In the Ising model, the magnetic field drives spins to prefer one direction over another and the coupling constant drives neighboring spins to align.  In the two-state system analogy considered here, the ``magnetic field'' drives the particle to favor one of the two potential wells, and the ``coupling constant'' recognizes that particle's state at time $t+\Delta t$ depends on its state at time $t$ and thus is responsible for the
temporal correlations of the system.  This analogy will be discussed further in Section \ref{section:MarkovRelation}.

The most convenient scheme for determining the partition function in Equation \ref{eqZ2} analytically
is to exploit periodic boundary conditions.  Though it is natural in the context of spins to talk about periodic boundary conditions, such conditions in our dynamical problem correspond nonsensically to time running in circles.  Nevertheless, the mathematical convenience of periodic boundary conditions makes them an attractive mathematical option.  (The errors incurred in making this simplifying assumption are discussed later.)  Under the assumption of periodic boundary conditions, the 1-D Ising model partition function can be evaluated analytically using transfer matrices as $Z=e^{-JN}(\lambda_{+}^N+\lambda_{-}^N)$ where $\lambda_{+}$ and $\lambda_{-}$ are the eigenvalues of the transfer matrix $T=\left( \begin{smallmatrix} e^{h+J}&e^{-J}\\ e^{-J}&e^{-h+J} \end{smallmatrix} \right) \label{eqMatrix}$ \cite{Pathria}.

As usual in maximum entropy problems, the values of the Lagrange multipliers are found by
evaluating partial derivatives of $Z$ \cite{Jaynes}.  In
particular, we need
\begin{equation}
 \frac{\partial \ln Z}{\partial h}=\langle i\rangle
 \label{eq:LM_h,i}
 \end{equation}
  and 
  \begin{equation}
  \frac{\partial\ln Z}{\partial J}=\frac{\partial\ln Z}{\partial\alpha}\frac{d\alpha}{dJ}=-2\langle N_s\rangle.
  \label{eq:LM_J,Ns}
  \end{equation}
Equations \ref{eq:LM_h,i} and \ref{eq:LM_J,Ns} are transcendental and can be solved numerically.
   
\subsection{Relation to MaxCal partition functional with Markov constraints}
\label{section:MarkovRelation}
   
Even though Equations \ref{eq:LM_h,i} and \ref{eq:LM_J,Ns} are transcendental, we can find closed-form expressions for $h$ and $J$ in terms of the rate constants $k_A$, $k_B$, and the time step $\Delta t$ by relating a previously used two-state MaxCal partition functional to the one used here.  In an earlier MaxCal analysis of the two-state system,  the constraints $N_{aa},~N_{ab},~N_{ba},$ and $N_{bb}$ were used instead of $N_s$ and $i$ \cite{Gerhard} \cite{Wuetal}.  The quantity $N_{xy}$ reports the number of time instants of length
$\Delta t$ out of a total of $N$ such time steps
where  the particle moved from state $x$ to state $y$.  The MaxCal partition function using this alternative set of constraints is
\begin{equation}
Z=\sum_{\Gamma}\gamma_{aa}^{N_{aa,\Gamma}}\gamma_{ab}^{N_{ab,\Gamma}}\gamma_{ba}^{N_{ba,\Gamma}}\gamma_{bb}^{N_{bb,\Gamma}},
\label{eqZold}
\end{equation}
where each $\Gamma$ is a different microtrajectory.

Note that the four constraints $\langle N_{aa}\rangle,~\langle N_{ab}\rangle,~\langle N_{ba}\rangle,$ and $\langle N_{bb}\rangle$ are not linearly independent, since the particle can only start in state A or B and move to state A or B in each of the $N$ time steps.   This constraint is captured by the relation
\begin{equation}
N_{\Gamma}=N_{aa,\Gamma}+N_{ab,\Gamma}+N_{ba,\Gamma}+N_{bb,\Gamma}.
\label{eq:constraintRelation1}
\end{equation}
To relate these constraints to $N_{s,\Gamma}$ and $i_{\Gamma}$, note that the total number of times that the particle switches states is the sum of the number of times it switches from state $A$ to state $B$ and vice versa,
\begin{equation}
N_{s,\Gamma}=N_{ab,\Gamma}+N_{ba,\Gamma}.
\end{equation}
Additionally, the average state of the particle is the weighted sum of the amount of time it spends in state $A$ and the amount of time it spends in state $B$,
\begin{equation}
i_{\Gamma}=\left(1\right)\left(N_{aa,\Gamma}+N_{ab,\Gamma}\right)+\left(-1\right)\left(N_{bb,\Gamma}+N_{ba,\Gamma}\right).
\end{equation}
Under the assumption of periodic boundary conditions used to calculate the partition function in Section \ref{section:IsingModel}, the particle must start and end in the same state, and therefore switch from state B to state A just as many times as it switches from state A to state B
resulting in the constraint
\begin{equation}
N_{ab,\Gamma}=N_{ba,\Gamma}.
\label{eq:constraintRelation4}
\end{equation}
The Equations \ref{eq:constraintRelation1}-\ref{eq:constraintRelation4} above allow us
to solve for the $N_{xy}$ quantities in terms of $N_s$ and $i$, namely, 
\begin{equation}
N_{aa,\Gamma}=\frac{N+i_{\Gamma}-N_{s,\Gamma}}{2},
\label{eqNaa}
\end{equation}
\begin{equation}
N_{bb,\Gamma}=\frac{N-i_{\Gamma}-N_{s,\Gamma}}{2},
\label{eqNbb}
\end{equation}
and
\begin{equation}
N_{ab,\Gamma}=N_{ba,\Gamma}=\frac{N_{s,\Gamma}}{2}.
\label{eqNab}
\end{equation}
If we substitute Equations \ref{eqNaa}, \ref{eqNbb}, and \ref{eqNab} into Equation \ref{eqZold}, we find that
\begin{eqnarray}
Z&=&\sum_{\Gamma} \gamma_{aa}^{N_{aa,\Gamma}}\gamma_{ab}^{N_{ab,\Gamma}}\gamma_{ba}^{N_{ba,\Gamma}}\gamma_{bb}^{N_{bb,\Gamma}} \\
&=&\exp(\frac{N\ln(\gamma_{aa}\gamma_{bb})}{2}) \times \sum_{\Gamma} \exp(\frac{\ln\gamma_{aa}}{2} (i_{\Gamma}-N_{s,\Gamma}) \nonumber \\
&& -\frac{\ln\gamma_{bb}}{2} (i_{\Gamma}+N_{s,\Gamma})+\frac{\ln(\gamma_{ba}\gamma_{ab})}{2} N_{s,\Gamma}) \\
&=& \exp(\frac{N\ln(\gamma_{aa}\gamma_{bb})}{2}) \times \nonumber \\
&& \sum_{\Gamma} \exp(\frac{i_{\Gamma}}{2}\ln(\frac{\gamma_{aa}}{\gamma_{bb}})+\frac{N_{s,\Gamma}}{2}\ln(\frac{\gamma_{ab}\gamma_{ba}}{\gamma_{aa}\gamma_{bb}})).
\label{eqZ3}
\end{eqnarray}
Therefore, the probability of the particle moving along a particular microtrajectory $\Gamma$ is
\begin{equation}
p_{\Gamma} = \frac{\exp(\frac{i}{2}\ln(\frac{\gamma_{aa}}{\gamma_{bb}})+\frac{N_s}{2}\ln(\frac{\gamma_{ab}\gamma_{ba}}{\gamma_{aa}\gamma_{bb}}))}{\sum_{\Gamma'} \exp(\frac{i_{\Gamma'}}{2}\ln(\frac{\gamma_{aa}}{\gamma_{bb}})+\frac{N_{s,\Gamma'}}{2}\ln(\frac{\gamma_{ab}\gamma_{ba}}{\gamma_{aa}\gamma_{bb}}))}. 
\end{equation}

To recapitulate, we have two different MaxCal formulations of the  two-state system dynamics, derived from two apparently different sets of constraints.  Their partition functions are given by Equation \ref{eqZ1} and Equation \ref{eqZ3}.  Both  models accurately describe the two-state system, as shown in Section \ref{section:Test} and Reference \cite{Wuetal}.   We can impose
the consistency condition that  both models have the same predictions for $p_{\Gamma}$, and hence they must have the same Lagrange multiplier coefficients for $N_{s,\Gamma}$ and $i_{\Gamma}$.  As a result, we can prescribe the equivalence of the two formulations through the relations
\begin{equation}
-2J=\frac{1}{2}\ln(\frac{\gamma_{ab}\gamma_{ba}}{\gamma_{aa}\gamma_{bb}})
\label{eqAlpha}
\end{equation}
and
\begin{equation}
h=\frac{1}{2}\ln(\frac{\gamma_{aa}}{\gamma_{bb}}).
\label{eqMu}
\end{equation}
Recall that these Lagrange multipliers are \cite{Wuetal}
\begin{equation}
\gamma_{ab}=k_A\Delta t,~\gamma_{ba}=k_B\Delta t,~\gamma_{aa}=1-k_A\Delta t,~\gamma_{bb}=1-k_B\Delta t.
\label{eq:MarkovLMsRates}
\end{equation} 
These statistical weights are Markovian in nature: $\gamma_{ab}$ is the probability that the particle will transition from state A to state B in the time interval $\Delta t$, $\gamma_{ba}$ is the probability that the particle will transition from state B to state A in the time interval $\Delta t$, and so on.  The alternative interpretation is that according to Equation \ref{eqAlpha}, the ``coupling constant'' increases as dwell times increase; Equation \ref{eqMu} suggests that the ``magnetic field'' increases as state A is favored over state B and vice versa.

Note that the calculations in \cite{Wuetal} used exact boundary conditions, whereas Equation \ref{eqZ2} exploited the analytic
convenience of  periodic boundary conditions.  As a result, any differences between the predictions of the two MaxCal partition functionals is an inheritance of the difference in boundary conditions.  In this way, we can measure to what extent potentially false assumptions about boundary conditions will affect the validity of moment predictions.  As we will see later, the particular choice of boundary conditions are practically irrelevant for the systems considered here because the particles are observed for many time steps.

\section{Testing the MaxCal model via numerical simulation}
\label{section:Test}

We wish to test our MaxCal model by comparing theoretical predictions to data obtained from simulations or experimental measurements of a two-state system.  In an earlier paper, we described the use of a dual-well optical trap that could be used to sculpt different two-state trajectories by tuning the properties of the adjacent wells \cite{Wuetal}.  In this paper, we continue along similar lines, but instead using simulated data for this system like that shown in Figure \ref{fig:Data}.  Trajectories were simulated according to the Langevin equation subject to an external potential \cite{GillespieSim},
\begin{equation}
\frac{dV(t)}{dt}=-\frac{\gamma}{M}V(t)+\frac{dU}{dx}+\frac{\sqrt{2\gamma k_B T}}{M}\Gamma(t)
\label{eqSimulation}
\end{equation}
where $U(x)$ is a potential of the form
\begin{equation}
U(x)=-E_{1}\exp(-\frac{(x-\mu_1)^2}{\sigma_1^2})-E_{2}\exp(-\frac{(x-\mu_2)^2}{\sigma_2^2}),
\end{equation}
meant to mimic the two-well landscape imposed by the optical traps.
The parameters $\mu_1$ and $\mu_2$ give the x-coordinates on which each well is centered; the parameters $\sigma_1$ and $\sigma_2$ refer to the width of each potential well; and $E_1$ and $E_2$ refer to the depth of each well.  In simulations, parameters $\mu_1$, $\mu_2$, $\sigma_1$, and $\sigma_2$ were kept constant.  However, fifteen different combinations of $E_1$ and $E_2$ were used to simulate fifteen different double potential wells.  See Table \ref{table:parameters} and Table \ref{table:parameters2} for a listing of $E_1$, $E_2$, and $\Delta E=E_1-E_2$ for each of the relevant energy landscapes.

\begin{table}[ht]
\centering
\begin{tabular}{c c}
\hline\hline
Parameter & Value \\ [0.5ex]
\hline
$\mu_1$ & $-270~nm$ \\
$\mu_2$ & $270~nm$ \\
$\sigma_1$ & $200~nm$ \\
$\sigma_2$ & $200~nm$ \\
$m$ & $0.55~pg$ \\
\hline
\end{tabular}
\caption{Parameters governing the geometry of the potential wells for
the dynamics simulation.  $\mu_i$ refers to the center of the $i^{th}$ well
and $\sigma_i$ is the RMS width of the potential.  The width and mean position of the well were the same for all simulated potential surfaces.  $m$ is the mass of the particle.}
\label{table:parameters}
\end{table}

\begin{table}[ht]
\centering
\begin{tabular}{c c c c}
\hline\hline
Potential & $\Delta E$ & $E_{1}$ & $E_{2}$ \\ [0.5ex]
Well \# & ($k_B T$) & ($k_B T$) & ($k_B T$) \\ [0.5ex]
\hline
1 & -2.5841 & 1.9009 & 4.4851 \\
2 & -2.2803 & 2.0607 & 4.3409 \\
3 & -1.9764 & 2.2243 & 4.2007 \\
4 & -1.6724 & 2.3918 & 4.0642 \\
5 & -1.3684 & 2.5629 & 3.9313 \\
6 & -1.0643 & 2.7375 & 3.8019 \\
7 & -0.7602 & 2.9156 & 3.6758 \\
8 & -0.4562 & 3.0969 & 3.5531 \\
9 & -0.1521 & 3.2814 & 3.4335 \\
10 & 0.1521 & 3.4690 & 3.3169 \\
11 & 0.4562 & 3.6595 & 3.2034 \\
12 & 0.7603 & 3.8529 & 3.0927 \\
13 & 1.0643 & 4.0492 & 2.9848 \\
14 & 1.3684 & 4.2481 & 2.8797 \\
15 & 1.6724 & 4.4496 & 2.7772 \\
\hline
\end{tabular}
\caption{Parameters governing the depths of the potential wells for the different
simulations.  The relative depth of the two wells varied between each simulated potential surface.}
\label{table:parameters2}
\end{table}

The simulations give us explicit realizations  of the trajectory distribution that we can compare to theoretical predictions of our MaxCal model via partial derivatives of the partition function.  For example, we can compute the fluctuations in the mean state as
\begin{eqnarray}
\frac{\partial^2\ln Z}{\partial h^2}&=&\frac{1}{Z}\frac{\partial^2 Z}{\partial h^2}-(\frac{\partial\ln Z}{\partial h})^2\\
&=&\sum_{\Gamma}p_{\Gamma}(\sum_{t}\sigma_{\Gamma}(t))^2-\langle i\rangle^2\\
&=&\langle i^2\rangle-\langle i\rangle^2.
\label{eqVarI}
\end{eqnarray}
Similarly, we can compute other correlation functions of interest such
as
\begin{equation}
\frac{\partial^2\ln Z}{\partial J\partial h}=-2(\langle N_s i\rangle-\langle N_s\rangle \langle i\rangle)
\label{eqCovariance}
\end{equation}
and
\begin{equation}
\frac{\partial^2\ln Z}{\partial J^2} = 4(\langle N_s^2\rangle-\langle N_s\rangle^2).
\label{eqVarNs}
\end{equation}

\begin{figure} 
  \includegraphics[width=8.6cm]{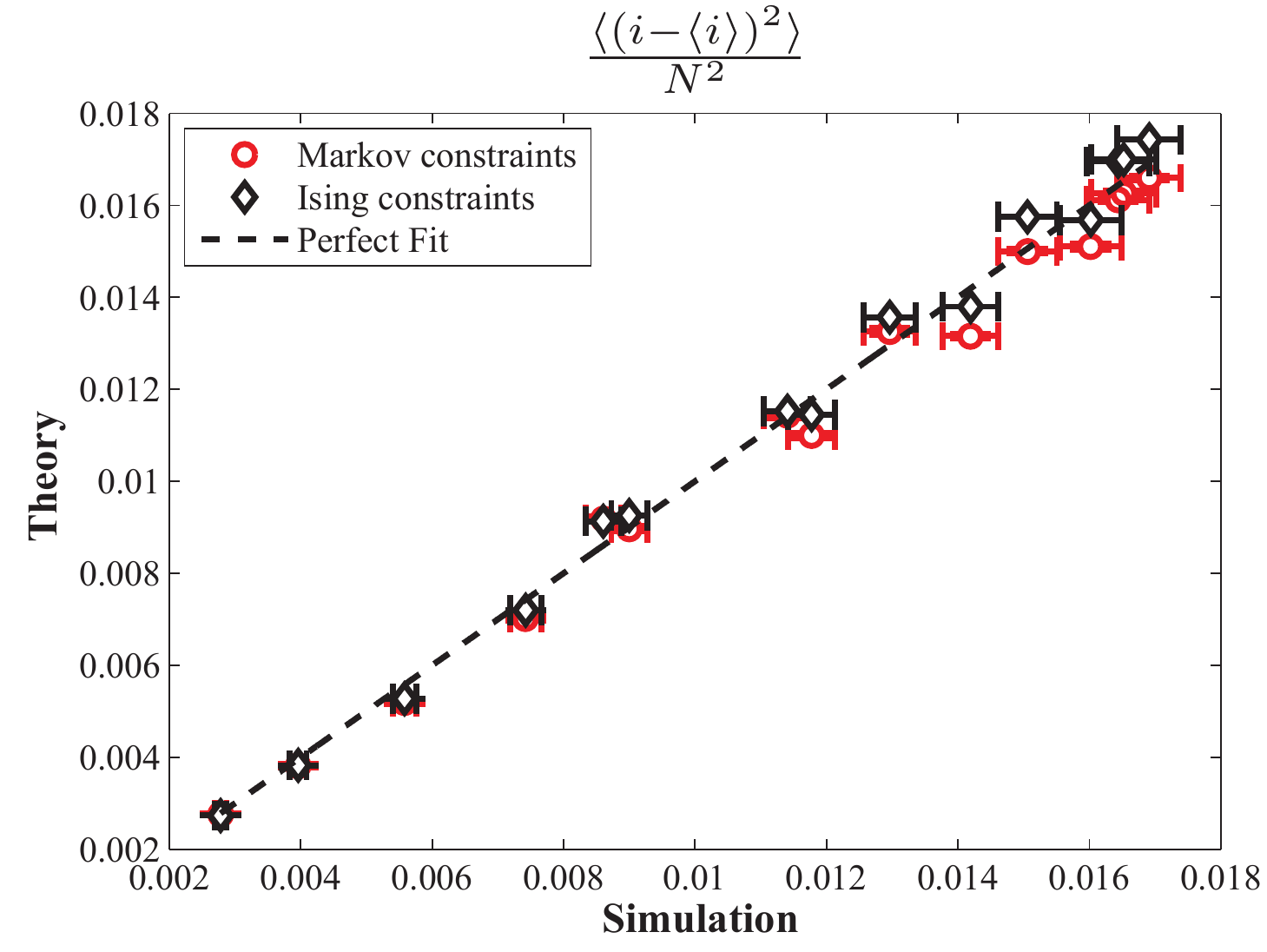}
  \caption{\textbf{Comparison between the value of the variance of $i$ obtained analytically and the variance obtained by ``measuring'' the properties of simulated trajectories.} 
  The graph shows the variance for each of the simulated potential
  surfaces characterized in Table~\ref{table:parameters2}, with each data point
  corresponding to a particular potential well.  The MaxCal results are shown
  for the cases when periodic boundary conditions and exact boundary
  conditions are used. The best fit line to the periodic boundary condition data is $y=1.034 x-2.5\times 10^{-4}$ with an $R^2=0.9945$; the best fit line to the exact boundary condition data is $y=0.9724 x-5.8\times 10^{-5}$ with an $R^2=0.9923$.}
  \label{fig:vari}
\end{figure}

\begin{figure}[h]
  \centering
  \includegraphics[width=8.6cm]{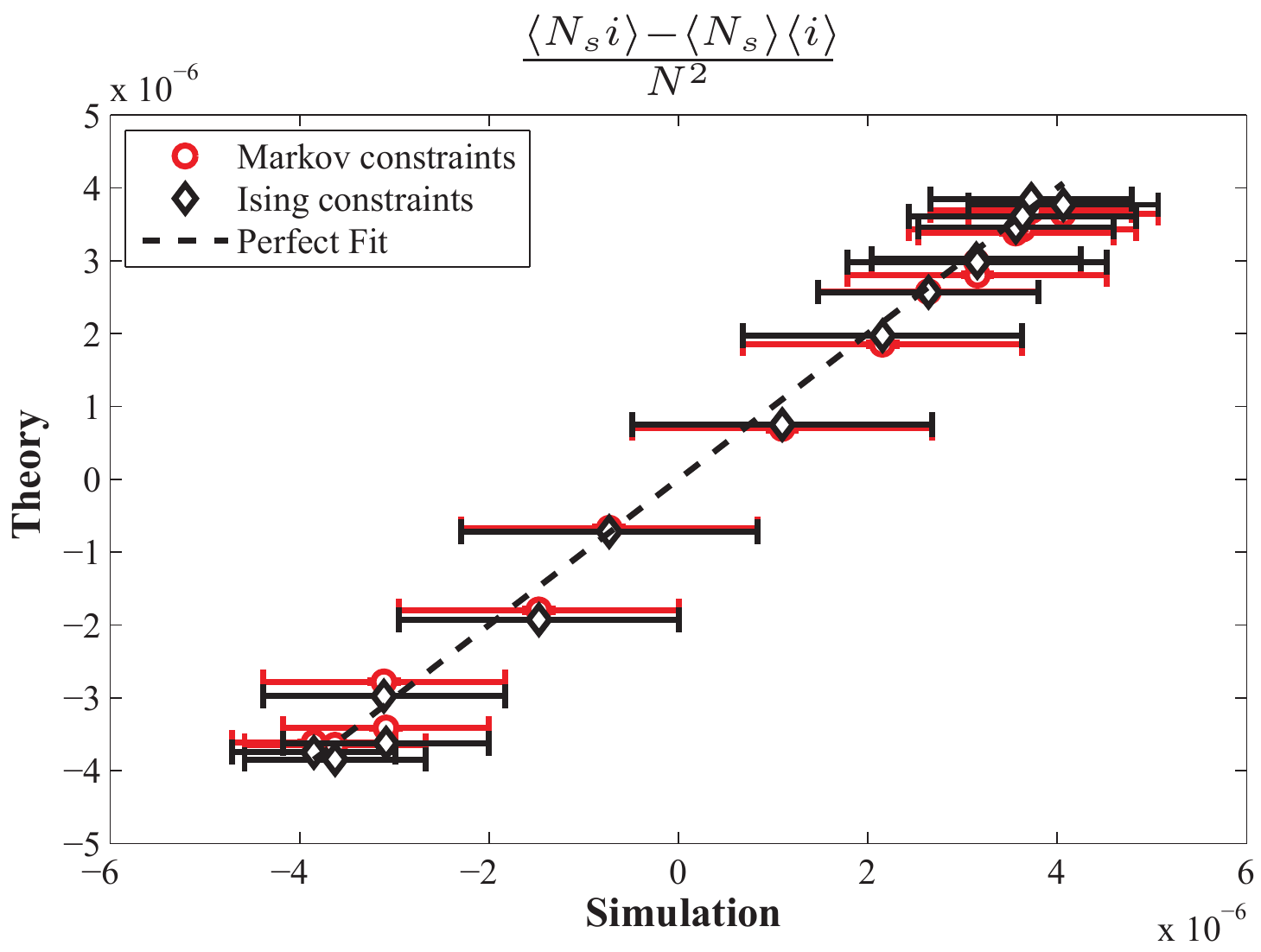}
  \caption{\textbf{Comparison between computed and simulated statistical
  correlations. }  
  The graph shows the statistical quantity labeled above the graph
   for each of the simulated potential
  surfaces characterized in Table~\ref{table:parameters2}, with each data point
  corresponding to a particular potential well.  The MaxCal results are shown
  for the cases when periodic boundary conditions and exact boundary
  conditions are used.  The best fit line to the periodic boundary condition data is $y=1.0063 x-1.5\times 10^{-7}$ with an $R^2=0.9957$; the best fit line to the exact boundary condition data is $y=0.9625 x-1.2\times 10^{-7}$ with an $R^2=0.9955$.}
  \label{fig:covariance}
\end{figure}

\begin{figure}[h]
	\centering
  \includegraphics[width=8.6cm]{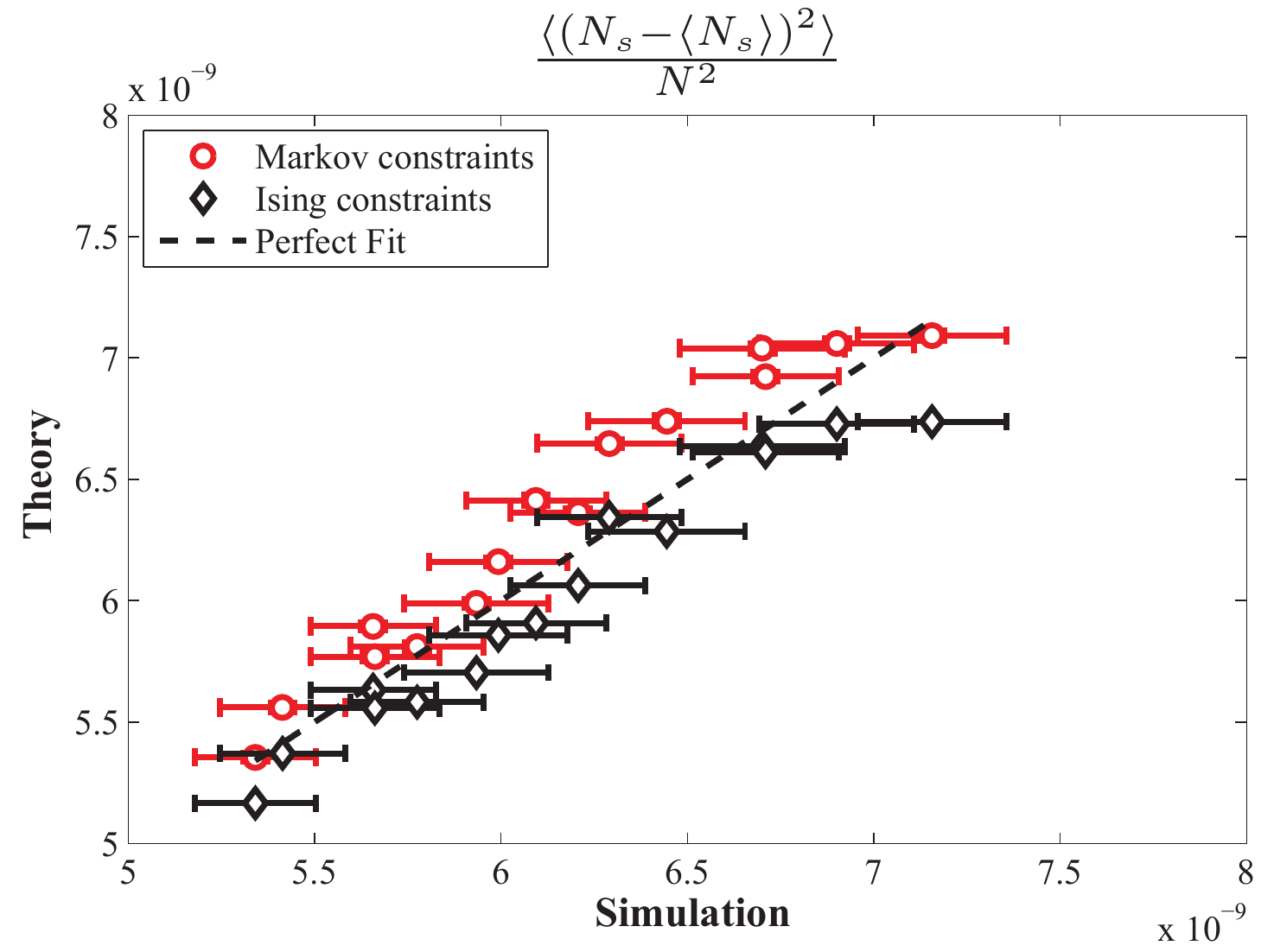}
  \caption{\textbf{Comparison between the value of the variance of $N_s$ obtained analytically and the variance obtained by ``measuring'' the properties of simulated trajectories.}   The graph shows the variance for each of the simulated potential
  surfaces characterized in Table~\ref{table:parameters2}, with each data point
  corresponding to a particular potential well.  The MaxCal results are shown
  for the cases of the partition functional used here (Equation \ref{eqZ1}) and of the Markov partition functional used in a previous analysis.  The best fit line to the periodic boundary condition data is $y=0.9297 x+2.9\times 10^{-10}$ with an $R^2=0.9626$; the best fit line to the exact boundary condition data is $y=1.0342 x-4.2\times 10^{-11}$ with an $R^2=0.9537$.}
  \label{fig:varNs}
  \end{figure}
  
To test our MaxCal model, we compared moments computed from the simulated data described previously with moments predicted by the MaxCal model  as
revealed in Equations \ref{eqVarI}, \ref{eqCovariance}, and \ref{eqVarNs}, for
example.  Figures \ref{fig:vari}, \ref{fig:covariance}, and \ref{fig:varNs} demonstrate good agreement between simulated data and the moments predicted from the MaxCal model via Equations \ref{eqVarI}, \ref{eqCovariance}, and \ref{eqVarNs}, respectively.  The calculations from this MaxCal model are also compared to the calculations from the MaxCal model described in Section \ref{section:MarkovRelation} \cite{Wuetal}.  In these graphs, each data point corresponds to a different potential well setup and error bars correspond to standard error.  Uncertainty in calculated values of moments stems from uncertainty in the value of the constraints taken from simulated data.  See Appendix I for details.

From Figures \ref{fig:vari}-\ref{fig:varNs}, there is good agreement between the two MaxCal models, which we have already argued are equivalent aside from a difference in boundary conditions.  This difference between the moment predictions in Figures \ref{fig:vari}-\ref{fig:varNs} is barely perceptible because $N=10^5>>1$, but there is certainly a significant difference in MaxCal predictions due to boundary conditions when there are very few timesteps \cite{Gerhard}.   For most cases of interest, the long-time limit will be the relevant situation in which the boundary conditions become less relevant as the total sampling time $N\Delta t$ increases compared to the time step $\Delta t$, just as surface terms become irrelevant in the thermodynamic limit in equilibrium statistical mechanics \cite{Pathria}.

\begin{figure}[h]
  \centering
  \includegraphics[width=8.6cm]{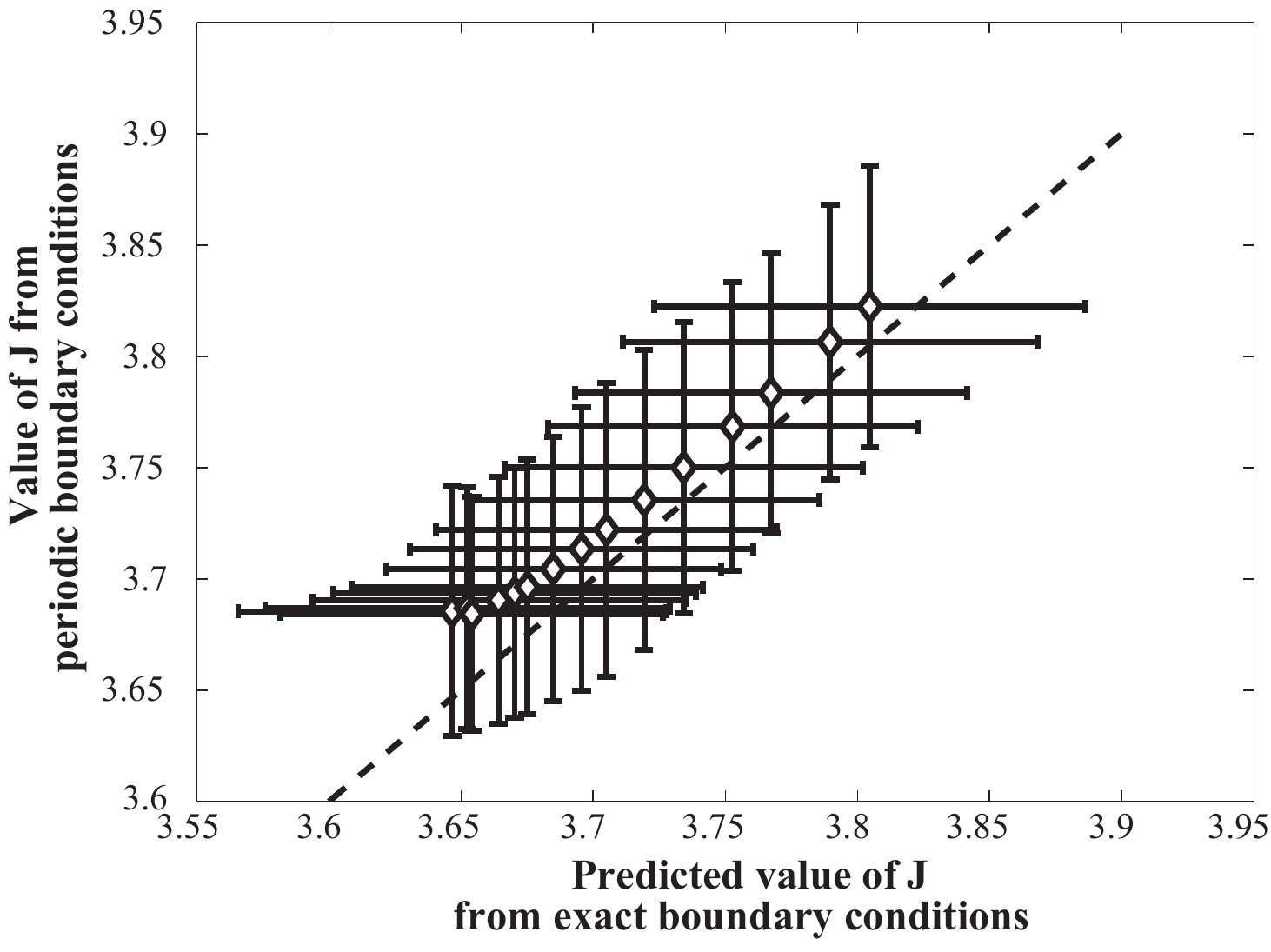}  
  \caption{\textbf{The Lagrange multiplier $J$.} Comparison between predicted value of $J$ from Equation (\ref{eqAlpha}) and its calculated value using MaxCal.  The dashed line shows where the data points would fall  if there were perfect agreement between the two MaxCal models.  The best fit line is given by $y=0.8944 x+0.4134$ with an $R^2=0.9884$.  Deviations from the proposed relationships in Equations \ref{eqAlpha} and \ref{eqMu} can be ascribed to the difference in boundary conditions, based on the analysis in Appendix II.}
  \label{fig:Alpha}
\end{figure}

\begin{figure}
  \includegraphics[width=8.6cm]{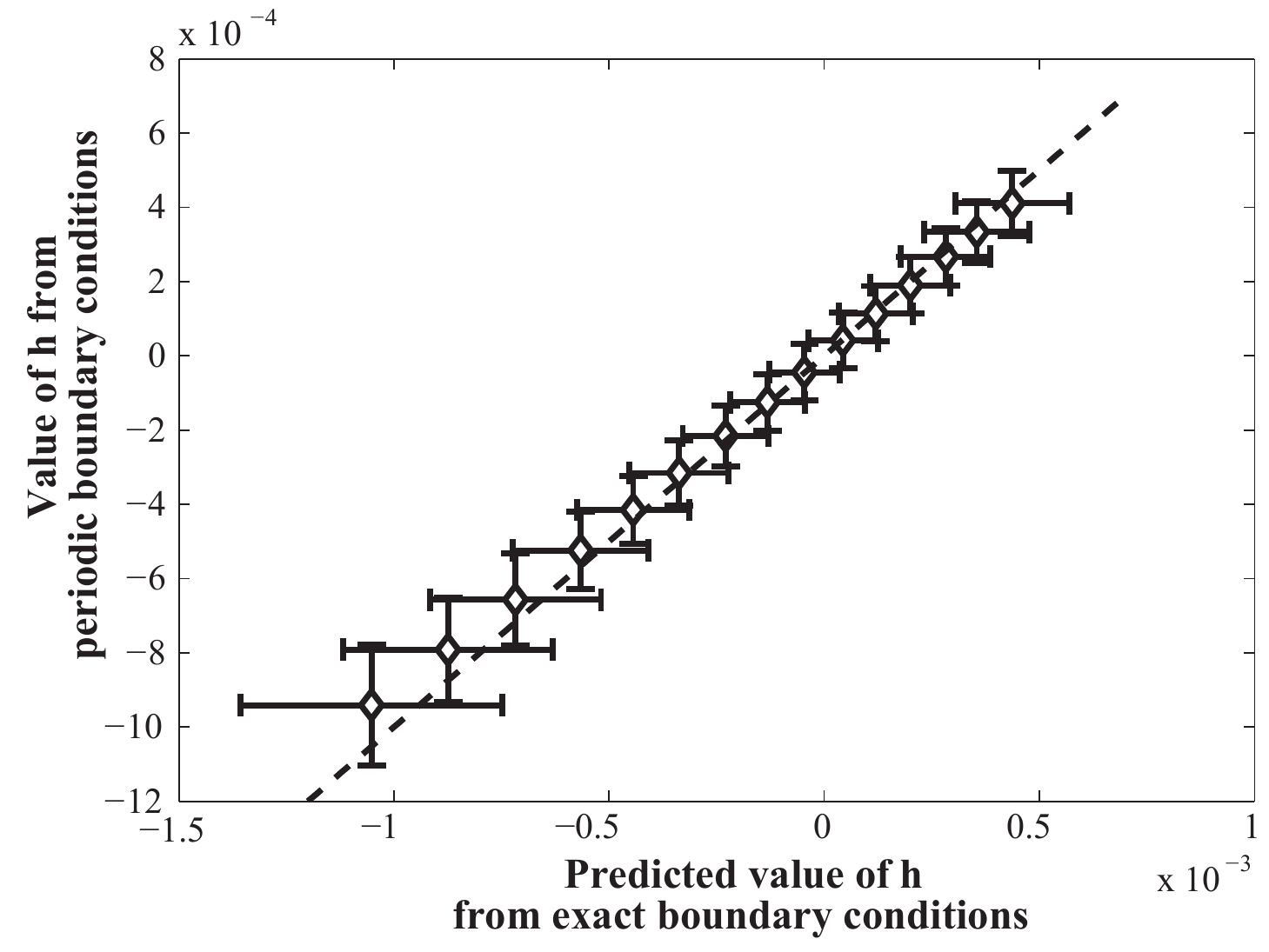}  
  \caption{\textbf{The Lagrange multiplier $h$.} Comparison between predicted value of $h$ from Equation (\ref{eqMu}) and its calculated values using MaxCal.  The dashed line shows where the data points would fall  if there were perfect agreement between the two MaxCal models.  The best fit line is given by $y=0.9154 x$ with an $R^2=0.9996$.}
  \label{fig:Mu}
\end{figure}

This indifference to boundary conditions can also be seen by testing Equations \ref{eqAlpha} and \ref{eqMu} derived for Lagrange multipliers in Section \ref{section:MarkovRelation}.  Recall that those equations hold only if periodic boundary conditions and exact boundary conditions lead to equivalent results, and therefore any deviation from the equations is due entirely to a difference in boundary conditions.  Figures \ref{fig:Alpha} and \ref{fig:Mu} demonstrate that Equations \ref{eqAlpha} and \ref{eqMu} agree well with the data from the fifteen different potential well setups described above, although Figure \ref{fig:Alpha} shows a systematic deviation due to a difference in boundary conditions.  We calculated rates $k_A$ and $k_B$ from Equation \ref{eq:MarkovLMsRates} and the values of the Lagrange multipliers derived using $\frac{\partial \ln Z}{\partial \ln \gamma_{aa}}=\langle N_{aa}\rangle$, $\frac{\partial \ln Z}{\partial \ln \gamma_{bb}}=\langle N_{bb}\rangle$, $\frac{\partial \ln Z}{\partial\ln \gamma_{ab}}=\langle N_{ab}\rangle$, and $\frac{\partial\ln Z}{\partial\ln \gamma_{ba}}=\langle N_{ba}\rangle$.  See Appendix II for more details.
%

The Ising model can also be used for predicting the autocorrelation function for the different states, so that MaxCal model calculations can be compared directly to predictions from the master equations for the two-state (i.e. random telegraph) problem \cite{Gardiner}.  In the language of the Ising model, the autocorrelation function for being in the state $A$ is
\begin{equation}
P(A,\tau)=\frac{1}{N}\langle\sum_t (\frac{\sigma(t)+1}{2})(\frac{\sigma(t+\tau)+1}{2})\rangle.
\end{equation}
The quantity $\frac{\sigma(t)+1}{2}$ is $1$ when the particle is in state A at time $t$ and $0$ if the particle is in state B; thus, the quantity $(\frac{\sigma(t)+1}{2})(\frac{\sigma(t+\tau)+1}{2})$ acts as a counter that is $1$ when the particle is in state A at times $t$ and $t+\tau$, and $0$ otherwise.  Expansion of this expression gives
\begin{equation}
P(A,\tau)=\frac{1}{4}(1+2\frac{\langle i\rangle}{N}+\langle\sigma(t)\sigma(t+\tau)\rangle).
\label{eqProbA}
\end{equation}
Similarly, we can write the corresponding autocorrelation function for the other state as
\begin{equation}
P(B,\tau)=\frac{1}{4}(1-2\frac{\langle i\rangle}{N}+\langle\sigma(t)\sigma(t+\tau)\rangle).
\label{eqProbB}
\end{equation}
Here $J,~h$ are not time dependent, so that $\langle\sigma(t)\sigma(t+\tau)\rangle$ is the same for any $t$.  We can evaluate $\langle\sigma(t)\sigma(t+\tau)\rangle$ by using standard transfer matrix techniques \cite{Pathria}, yielding the result
\begin{eqnarray}
\langle\sigma(t)\sigma(t+\tau)\rangle&=&\sum_{\Gamma}\sigma(t)\sigma(t+\tau)p_{\Gamma}\\
&=&\frac{1}{Z}tr(D^t(S^{-1}PS)D^{\tau}(S^{-1} \nonumber \\
&&PS)D^{N-t-\tau})
\end{eqnarray}
where $T$ is given by Equation (\ref{eqMatrix}) and $P$ is the Pauli spin matrix $\left( \begin{smallmatrix} 1&0\\ 0&-1 \end{smallmatrix} \right)$.  We find that
\begin{equation}
S^{-1} P S= \left( \begin{smallmatrix} x&-x-1\\ x-1&-x \end{smallmatrix} \right) ,
\end{equation}
where 
\begin{equation}
x=\frac{e^{2J}(e^{2h}-1)}{\sqrt{4e^{2h}+e^{4J}(e^{2h}-1)^2}}. 
\end{equation}
 Recall that $D=\left( \begin{smallmatrix} \lambda_{+}&0\\ 0&\lambda_{-} \end{smallmatrix} \right)$, which makes exponentiation trivial, resulting in
\begin{equation}
D^n=\left( \begin{smallmatrix} \lambda_{+}^n&0\\ 0&\lambda_{-}^n \end{smallmatrix} \right).
\end{equation}
Therefore, matrix multiplication gives
\begin{eqnarray}
\langle\sigma(t)\sigma(t+\tau)\rangle&=&\frac{e^{4J}(e^{2h}-1)^2}{4e^{2h}+e^{4J}(e^{2h}-1)^2}- \nonumber \\
&&\frac{4e^{2h}}{4e^{2h}+e^{4J}(e^{2h}-1)^2}\times \nonumber \\
&&\frac{\lambda_{+}^\tau \lambda_{-}^{N-\tau}+\lambda_{-}^\tau \lambda_{+}^{N-\tau}}{\lambda_{+}^N+\lambda_{-}^N},
\end{eqnarray}
which we can use to directly calculate $P(A,t)$ and $P(B,t)$.

\begin{figure}[htbp]
  \begin{center}%
  \includegraphics[width=8.6cm]{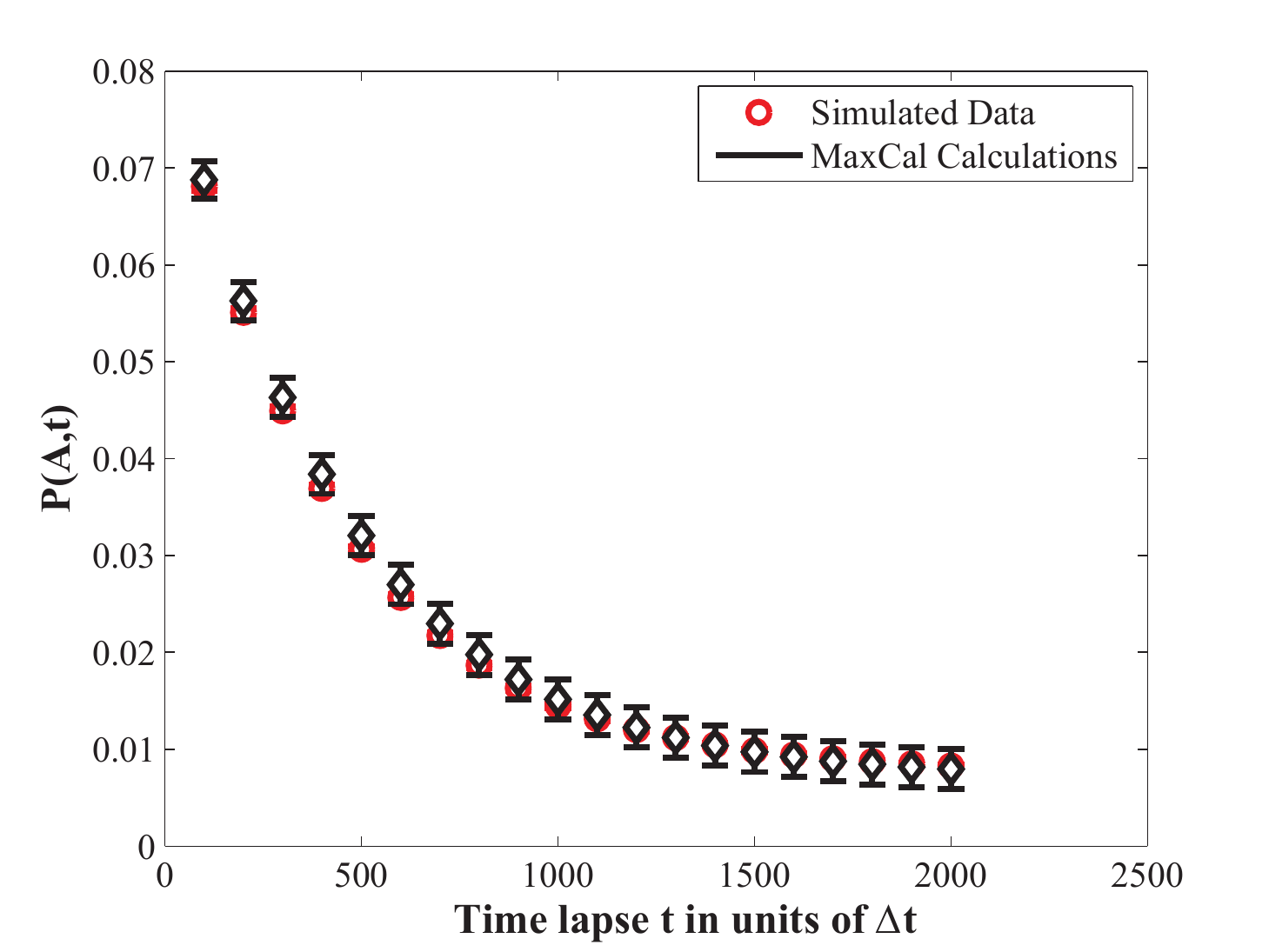}    
  \end{center}
\caption{\textbf{Time dependence of the correlation function.} The graph of Equation \ref{eqProbA} is shown as a function of time in units of $\Delta t$ for potential well number one.}
  \label{fig:Correlation}
\end{figure}

A comparison between simulated and theoretical values for $P(A,t)$ is shown in Figure \ref{fig:Correlation}.  Equations \ref{eqProbA} and \ref{eqProbB} are quite accurate, although they are not identical to the probabilities that one would calculate using master equations.  The reason for this is that our discrete MaxCal model has a third implicit parameter in addition to $k_A$ and $k_B$: $\Delta t$, the time scale associated with the discretization of the trajectories.  The exact nature of the dependence on $\Delta t$ was made explicit in Section \ref{section:MarkovRelation}.

There are benefits and drawbacks to using the MaxCal analysis presented here instead of a more traditional master equation analysis.  For example, it is straightforward to calculate the joint probability distribution of $N_s$ and $i$, $p(N_s,i)$, using the  MaxCal model.  Indeed,
for some quantities we have not even been able to figure out how to compute them
using the master equation treatment.  The ease of these calculations in
the MaxCal setting results from the fact that the moments of the probability distribution $p(N_s,i)$ are simply partial derivatives of the partition function $Z$ in our MaxCal model, and in the master equation formalism, one must calculate $p(N_s,i)$ from $P(A,t)$ and $P(B,t)$.
It is straightforward to calculate autocorrelations using the master equation formalism via a linear differential equation; the same quantity can be calculated from MaxCal using transfer matrices.

\section{Maxwell-like relations for dynamic processes}

Both in classical equilibrium thermodynamics and statistical mechanics
and in the thermodynamics of small departures from equilibrium, there are
broad classes of relations such as the Maxwell relations and the Onsager
relations that illustrate the mathematical linkage of quantities that at first
blush might seem unrelated.    In some cases, the simplest way to 
explain these relations is that they reflect equality of certain second-order
mixed partial derivatives.     The maximum caliber formalism admits similar
relations \cite{Jaynes2,Gerhard}.   In the language of trajectories, what these relations tell us is
that if we have measured properties such as 
$N_s$ and $i$ on one two-well landscape, we can predict what the dynamics
will be like on a ``nearby'' landscape without doing any further
measurements.

In this MaxCal model, Maxwell-like relations fall out as the equality of mixed partial derivatives of $\ln Z$.  For example, the statement that $\frac{\partial^2\ln Z}{\partial h\partial J}=\frac{\partial^2\ln Z}{\partial J\partial h}$ leads to
\begin{equation}
\frac{\partial \langle N_s\rangle}{\partial h}=-\frac{1}{2}\frac{\partial^2\ln Z}{\partial h\partial J}=-\frac{1}{2}\frac{\partial \langle i\rangle}{\partial J}.
\label{eqOnsager}
\end{equation}
Other results can be obtained from higher-order mixed partial derivatives, e.g. $\frac{\partial^3\ln Z}{\partial J\partial^2 h}=\frac{\partial^2}{\partial J\partial h}(\frac{\partial\ln Z}{\partial h})$ yields $-\frac{1}{2}\frac{\partial^2 \langle N_s\rangle}{\partial h^2}=\frac{\partial^2\langle i\rangle}{\partial h^2}$. It still remains to see how to turn such relations into general useful
predictive tools for the design and interpretation of experiments.

\section{Conclusion}

In summary, we have studied a single colloidal particle undergoing a two-state process with stationary rates, though the problem with time-varying rates is of great interest.  This classic problem is studied using the theory of maximum caliber.   By measuring relatively short trajectories ($\sim$ one hour), we are able to find the full probability distribution of trajectories from a MaxCal model using the average frequency of state-switching and the average state as constraints, and confirm that the predicted trajectory distribution agrees with simulated data.  Additionally, we show that the ``Markov-like'' constraints used in an earlier MaxCal analysis of the two-state system \cite{Wuetal} differs from the MaxCal analysis presented here only via a difference in boundary conditions.  The mapping of the Ising-like MaxCal model onto the Markov-like MaxCal model allows us to assign a physical interpretation to the Lagrange multipliers.

Mapping two-state dynamics onto a one-dimensional Ising model yields unexpected insights into two-state dynamics, even though the master equations that describe two-state kinetics have been used for decades.  First, we show that the dynamics of the two-state system can be mapped onto a 1-D Ising model whose ``coupling constant'' and ``magnetic field'' can be written explicitly in terms of $k_A$, $k_B$, and $\Delta t$.  Then we derive several relationships that hold true specifically for discretized two-state systems, i.e. the expressions depend on the time step $\Delta t$ as well as the rates $k_A$ and $k_B$.  We derive and verify expressions for the moments of the joint probability distribution $p(N_s,i)$, which  to the best of our knowledge have not been derived previously using the  traditional master equation formalism or anything else.  Next we derive and verify an expression for the autocorrelation functions $P(A,t)$ and $P(B,t)$ in a discretized system.   It is possible that these new formulas will be of practical value to those who study systems that can be modeled as two-state systems, e.g. ion channels and ligand-receptor complexes.

\section{Acknowledgements}

We are grateful to Ken Dill and Kings Ghosh for years of
discussion on the MaxCal formalism.  SM acknowledges the support of a Caltech Student Undergraduate Research Fellowship.  DW acknowledges the
support of a NIH UCLA-Caltech MD-PhD fellowship.
This work was also supported by the NIH Director's Pioneer award.

\section*{Appendix I}

Because of the finite time scale of our simulations,
there is uncertainty in  our simulation measurements of $\langle i\rangle$ and $\langle N_s\rangle$;  there is a standard error associated with both measurements-- $\langle i\rangle\pm\sigma_i$ and $\langle N_s\rangle\pm\sigma_{N_s}$.  Recall that the Lagrange multipliers $J$ and $h$ are determined by solving transcendental equations $\frac{\partial\ln Z}{\partial J}=\langle 2N_s\rangle$ and $\frac{\partial\ln Z}{\partial h}=\langle i\rangle$.  These values must be modified in order to account for standard deviation $\sigma_{N_s}$ and $\sigma_i$ in our estimation of constraints $\langle N_s\rangle$ and $\langle i\rangle$.
We can determine the standard deviation of Lagrange multipliers, $\sigma_{J}$ and $\sigma_{h}$, by using error propagation on $\langle i\rangle$ and $\langle N_s\rangle$, namely,
\begin{equation}
\sigma_{i}^2=(\frac{\partial\langle i\rangle}{\partial h}\sigma_{h})^2+(\frac{\partial\langle i\rangle}{\partial J}\sigma_{J})^2
\label{eqSigmaI}
\end{equation}
and
\begin{equation}
\sigma_{N_s}^2=(\frac{\partial\langle N_s\rangle}{\partial h}\sigma_{h})^2+(\frac{\partial\langle N_s\rangle}{\partial J}\sigma_{J})^2.
\label{eqSigmaNs}
\end{equation}
The error bars in Figures \ref{fig:Alpha} and \ref{fig:Mu} are the standard devations from Equations \ref{eqSigmaI} and \ref{eqSigmaNs}.

We can then use Equations (\ref{eqSigmaI}) and (\ref{eqSigmaNs}) to solve for $\sigma_{h}$ and $\sigma_{J}$.  Finding the standard deviation of second moments follows
from a similar strategy in error propagation.  For instance, $X=\langle i^2\rangle-\langle i\rangle^2=\frac{\partial^2\ln Z}{\partial h^2}$.  The standard deviation of this quantity is
\begin{eqnarray}
\sigma_{X}^2&=&(\frac{\partial X}{\partial h})^2\sigma_{h}^2+(\frac{\partial X}{\partial J})^2\sigma_{J}^2 \\
&=& (\frac{\partial^3\ln Z}{\partial h^3})^2\sigma_{h}^2+(\frac{\partial^3\ln Z}{\partial J\partial h^2})^2\sigma_{J}^2.
\end{eqnarray}
In this equation, $\frac{\partial^3\ln Z}{\partial h^3}$ and $\frac{\partial^3\ln Z}{\partial J\partial h^2}$ can be evaluated at $J$ and $h$, and $\sigma_{J}$ and $\sigma_{h}$ are obtained via Equations (\ref{eqSigmaI}) and (\ref{eqSigmaNs}).  Thus the standard deviation of $\langle i^2\rangle-\langle i\rangle^2$ can be evaluated.
This analysis extends to the other second moments.  Let $Y=\langle N_s i\rangle -\langle N_s\rangle\langle i\rangle=-\frac{1}{2}\frac{\partial^2\ln Z}{\partial J\partial h}$ and $Z=\langle N_s^2\rangle-\langle N_s\rangle^2=-\frac{1}{2}\frac{\partial^2\ln Z}{\partial J^2}$.  Similar analysis to that above shows:
\begin{equation}
\sigma_{Y}^2=(-\frac{1}{2}\frac{\partial^3\ln Z}{\partial h^2\partial J})^2\sigma_{h}^2+(-\frac{1}{2}(\frac{\partial^3\ln Z}{\partial J^2\partial h})^2\sigma_{J}^2
\end{equation}
and
\begin{equation}
\sigma_{Z}^2=(-\frac{1}{2}\frac{\partial^3\ln Z}{\partial h\partial J^2})^2\sigma_{h}^2+(-\frac{1}{2}(\frac{\partial^3\ln Z}{\partial J^3})^2\sigma_{J}^2.
\end{equation}
What is plotted in Figures \ref{fig:vari}-\ref{fig:varNs} and Figure \ref{fig:Correlation} is the standard error, which is the standard deviation divided by $\sqrt{M}$, where $M$ is the number of trajectories simulated.  For each different potential surface, $2200$ trajectories were simulated, and so $M=2200$.

\section*{Appendix II}

Previously in Section \ref{section:Test} we saw a systematic underestimation of the measured value of $J$.  To understand this, we would like to compare values of $J$ calculated using the same constraints $\langle i\rangle,~\langle N_s\rangle$ and same MaxCal model in Equation \ref{eqZ2}, just with three different types of boundary conditions-- exact, periodic, and free.  Even though we cannot analytically evaluate the partition function under free boundary conditions, we can still find the value of $J$ under free boundary conditions following methods in earlier work \cite{InamdarThesis}.  Under free boundary conditions,
\begin{equation}
p(N_s)={N-1 \choose N_s}\frac{e^{-2J N_s}}{(1+e^{-2J})^{N-1}}.
\end{equation}
Thus, we can evaluate the Lagrange multiplier $J$ as a function of the constraint $\langle N_s\rangle$ using
\begin{equation}
\langle N_s\rangle = \sum_{N_s=0}^{N-1} N_s p(N_s),
\end{equation}
which, after some straightforward algebra, yields
\begin{equation}
J=-\frac{1}{4}\ln(\frac{N-1}{\langle N_s\rangle}-1).
\label{eqnLMFree}
\end{equation}
To recap, we now have three different values for $J$ depending on boundary conditions used to evaluate the same MaxCal partition function in Equation \ref{eqZ2}.  Under periodic boundary conditions, $J$ is found from numerically solving the closed form, transcendental system of equations given by Equations \ref{eq:LM_h,i} and \ref{eq:LM_J,Ns}.  Under exact boundary conditions, $J$ is given by Equation \ref{eqAlpha}.  Under free boundary conditions, $J$ is given by Equation \ref{eqnLMFree}.  Our goal is to understand whether or not boundary conditions is really driving the difference between the three different $J$'s by examining how the difference between the different $J$'s varies with the constraint $\langle i\rangle$.

\begin{figure}[h]
	\centering
	\includegraphics[width=8.6cm]{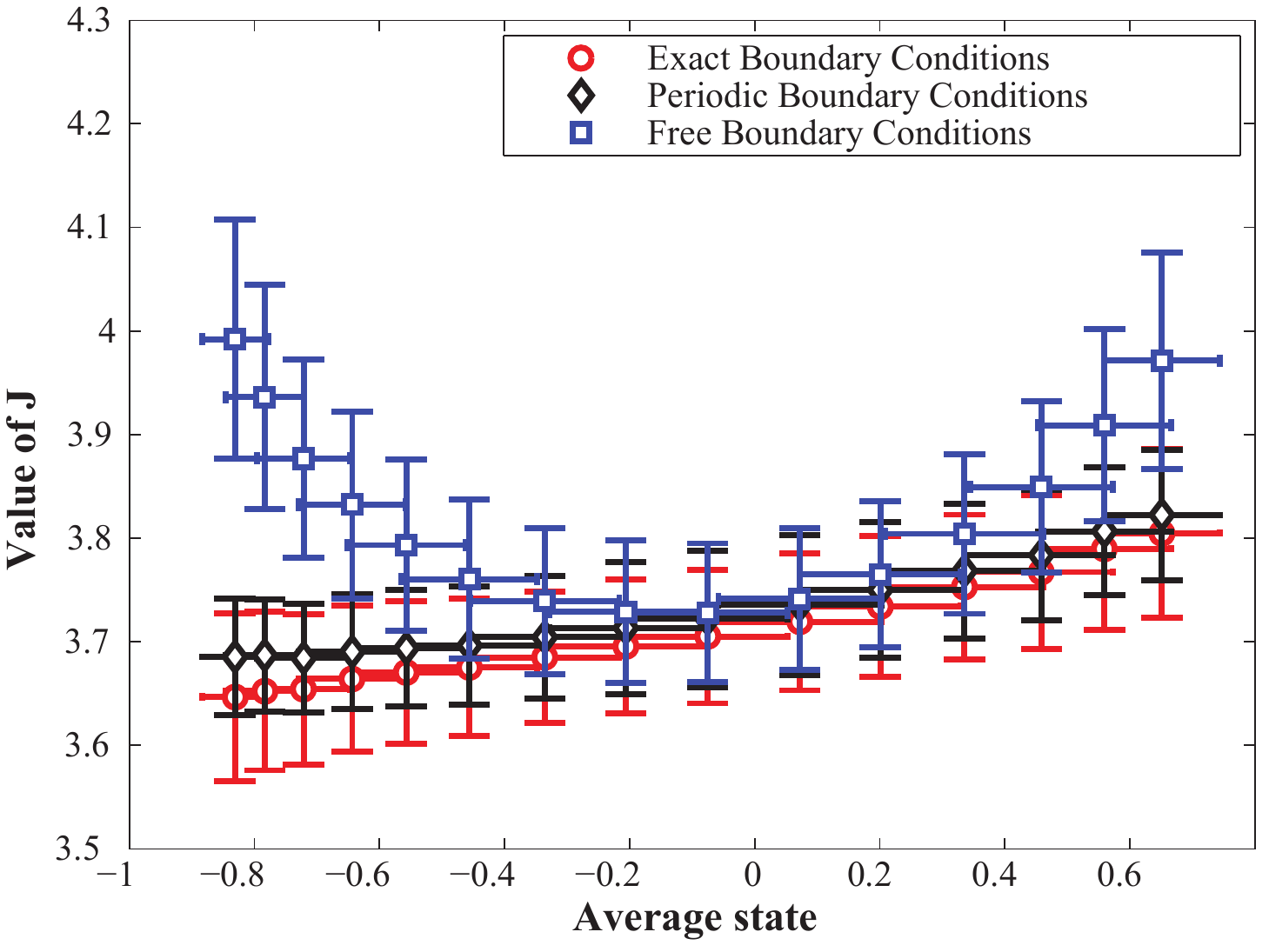}
	\caption{\textbf{Boundary conditions cause systematic deviations in evaluations of the Lagrange multiplier $J$.} The x-axis shows the absolute value of the average state $|\frac{\langle i\rangle}{N}|$, which is a measure of how biased the particle is towards being in either state.  The y-axis shows $J$ from: the predicted value of $J$ from Equation \ref{eqAlpha}, its calculated value using MaxCal, or its calculated value assuming free boundary conditions in Equation \ref{eqnLMFree}.  The data show that the difference in calculated values of $J$ between periodic/exact boundary conditions and free boundary conditions increases with the driving force to be in a particular state.  See text for further explanation.}
	\label{fig:Alpha2}
\end{figure}

The average state $\frac{\langle i\rangle}{N}$ is a sort of metric for how hard the particle is pushed to be in one state or the other.  Periodic boundary conditions become more like exact boundary conditions as the driving force to be in either state increases, since then the probability that a trajectory will start and end in the same state increases.  Free periodic boundary conditions become more like exact boundary conditions as $|\frac{\langle i\rangle}{N}|$ approaches $0$, since when neither state is preferred, the particle is just as likely to start and end in either state.  In Figure \ref{fig:Alpha2}, as the average state approaches $1$ and $-1$, exact and periodic boundary conditions split away from free periodic boundary condition estimates. For $\langle i \rangle$ close to $0$, the three boundary conditions are very close together.  This is what we would expect if boundary conditions were indeed driving the difference between the different Lagrange multiplier values.  

\end{document}